\documentclass[reprint,nofootinbib,prd]{revtex4-2}
\usepackage[utf8]{inputenc} 
\usepackage{graphicx} 
\usepackage{dcolumn}
\usepackage{bm}
\usepackage{multirow}
\usepackage{amsmath}
\usepackage{amssymb}
\usepackage{tensor}

\newcommand{\rg}{\rightarrow}
\newcommand{\A}{1-\frac{2}{r}}
\newcommand{\AR}{1-\frac{2}{R}}
\newcommand{\li}{\ell}
\newcommand{\lp}{\left(}
\newcommand{\rp}{\right)}

\begin{document}

\title{Extreme mass-ratio binary black hole merger: \\ Characteristics of the test-particle limit}
\author{Barak Rom}
    \email{barak.rom@mail.huji.ac.il}
\author{Re'em Sari}
\affiliation{Racah Institute of Physics, The Hebrew University of Jerusalem, 9190401, Israel}

\date{\today} 

\begin{abstract}
We study binary black hole mergers in the extreme mass-ratio limit. We determine the energy, angular momentum, and linear momentum of the post-merger, remnant black hole. Unlike previous works, we perform our analysis directly in the test-particle limit by solving the Regge-Wheeler-Zerilli wave equation with a source that moves along a geodesic. We rely on the fact that toward the merger, small mass-ratio binary systems follow a quasiuniversal geodesic trajectory. This formalism captures the final premerger stages of small mass-ratio binaries and thus provides a straightforward universal description in a region inaccessible to numerical relativity simulations. We present a general waveform template that may be used in the search for gravitational wave bursts from small and intermediate mass-ratio binary systems. Finally, this formalism gives a formal proof that the recoil velocity is quadratic in the symmetric mass ratio $\nu$. Specifically, the velocity is given by $V/c\approx 0.0467 \nu^2$. This result is about $4\%$ larger than previously estimated. Most of this difference stems from the inclusion of higher multipoles in our calculation.
\end{abstract}

\maketitle

\section{Introduction}
The recent detections of gravitational waves (GW) \cite{GWTC} are the result of a decades-long theoretical and technological endeavor. The numerical relativity (NR) breakthrough \cite{NR1,NR2,NR3} promoted the extensive, ongoing research of comparable mass-ratio binary black hole mergers, which are the primary GW sources of current earth-based GW detectors. The study of small mass-ratio binary mergers, was initiated by the pioneering works of Regge, Wheeler, and Zerilli \cite{RW,Zer}, who formulated a simple wave equation from the field equations of general relativity, given the Schwarzschild geometry. Teukolsky \cite{Tek}, using Newman-Penrose formalism, generalized this to Kerr geometry as well.

Since then, substantial progress has been done in the study of the small mass-ratio binary merger, its corresponding GW emission \cite{B04,H06,N07,DN07,N10,H10,N11,BH11,K11,N14b,BH14,NR4,NR5}, and recoil velocity \cite{H04,B05,D06,S06,L07,S07,H10,N10,N11,K11,N14}. Moreover, the ability to produce fast and accurate waveform templates \cite{H07,BH12,G15,ND18,K20,H21a,H21b,TC22,H22} will play an essential role in the next generation GW observatories, as the Laser Interferometer Space Antenna (LISA), for which small mass-ratio binaries will be a prominent GW source.

In this paper, we investigate the merger of a nonspinning binary black hole (BH) system, initially on a quasicircular orbit, in the extreme mass-ratio limit; $\mu\ll M$, where $M$ is the total mass, and $\mu$ is the reduced mass of the system. The merger scenario can be qualitatively divided into three stages: ($i$) Quasicircular inspiral, during which the orbit evolves by the emission of GW. ($ii$) Universal plunge, where the infall path tends to the geodesic universal infall (GUI) trajectory. ($iii$) Quasinormal modes (QNM) ringdown.

A binary system, initially at large separation, emits GW which leads to a fast circularization of the orbit followed by a slow decrease of its semimajor axis \cite{P64}, with a radial velocity that is much smaller than the angular one. This is the quasicircular inspiral stage, which continues until the secondary BH crosses the innermost stable circular orbit (ISCO), at $R_{ISCO}=6M$. External to the ISCO, there are stable circular orbits, so the secondary BH slowly descends from one to another, but after the ISCO, there are no further stable circular orbits. Therefore, the secondary BH motion smoothly shifts from the GW-driven inspiral to a geodesic free fall. Finally, after the secondary BH crosses the peak of the curvature potential, at $R\sim3M$, the GW signal is predominated by the QNM ringing, the intrinsic vibration modes of the remnant BH \cite{QNM09}.

The initial motivation for this work lies in the insight that the plunge path, of systems that approach the merger along quasicircular orbits, is universal in the sense that it is insensitive to the initial separation and the exact mass ratio. Although the mass ratio determines the number of orbits internal to the ISCO, for any small mass ratio, the infall path tends to the GUI trajectory; namely, it coincides with the free fall trajectory of a test particle that is initially at the ISCO. Therefore, phenomena that strongly depend on the final premerger orbits, like the recoil velocity, are quasiuniversal.

We determine the postmerger energy, angular momentum, and linear momentum of the remnant BH and present the merger waveform, at the test-particle limit. The energy and angular momentum, up to first order in the mass ratio, are derived from their values at the ISCO, while the recoil velocity is numerically calculated by solving the RWZ equation for a source that moves along the GUI trajectory. This method allows for calculating directly at the test-particle limit, without introducing any finite mass ratio and extrapolating to the $\mu\rg0$ limit. For a small, finite mass ratio, there will be higher order corrections beyond the first order value, which we calculate in this paper. Finally, we sum the high multipoles contribution and thus evaluate the full recoil coefficient, by extrapolating the results to higher $\li$ than we numerically calculate.

Throughout the paper we use geometric units, $G=c=1$. 

\section{Relativistic Dynamics}
The equations of motion of a test particle that moves around a nonspinning BH, as derived from the Schwarzschild metric, are:
\begin{subequations}
\begin{gather}
\frac{d\Phi}{dt} =\lp1-\frac{2}{R}\rp\frac{L}{ER^2}\label{eq:b4}\\
\frac{dR}{dt} =\pm\frac{\left(1-\frac{2}{R}\right)}{E}\sqrt{E^2-\left(1-\frac{2}{R}\right)\left(1+\frac{L^2}{R^2}\right)},\label{eq:bb4}
\end{gather}
\end{subequations} 
where $E$ and $L$ are the energy and angular momentum, per unit mass, respectively, and $(t,R,\Phi)$ are the Schwarzschild coordinates. Note that $t$ and $R$ are measured in unites of $M$.

\subsection{The GUI trajectory}
\begin{figure}[ht!]
\centering
\includegraphics[width=8.6cm]{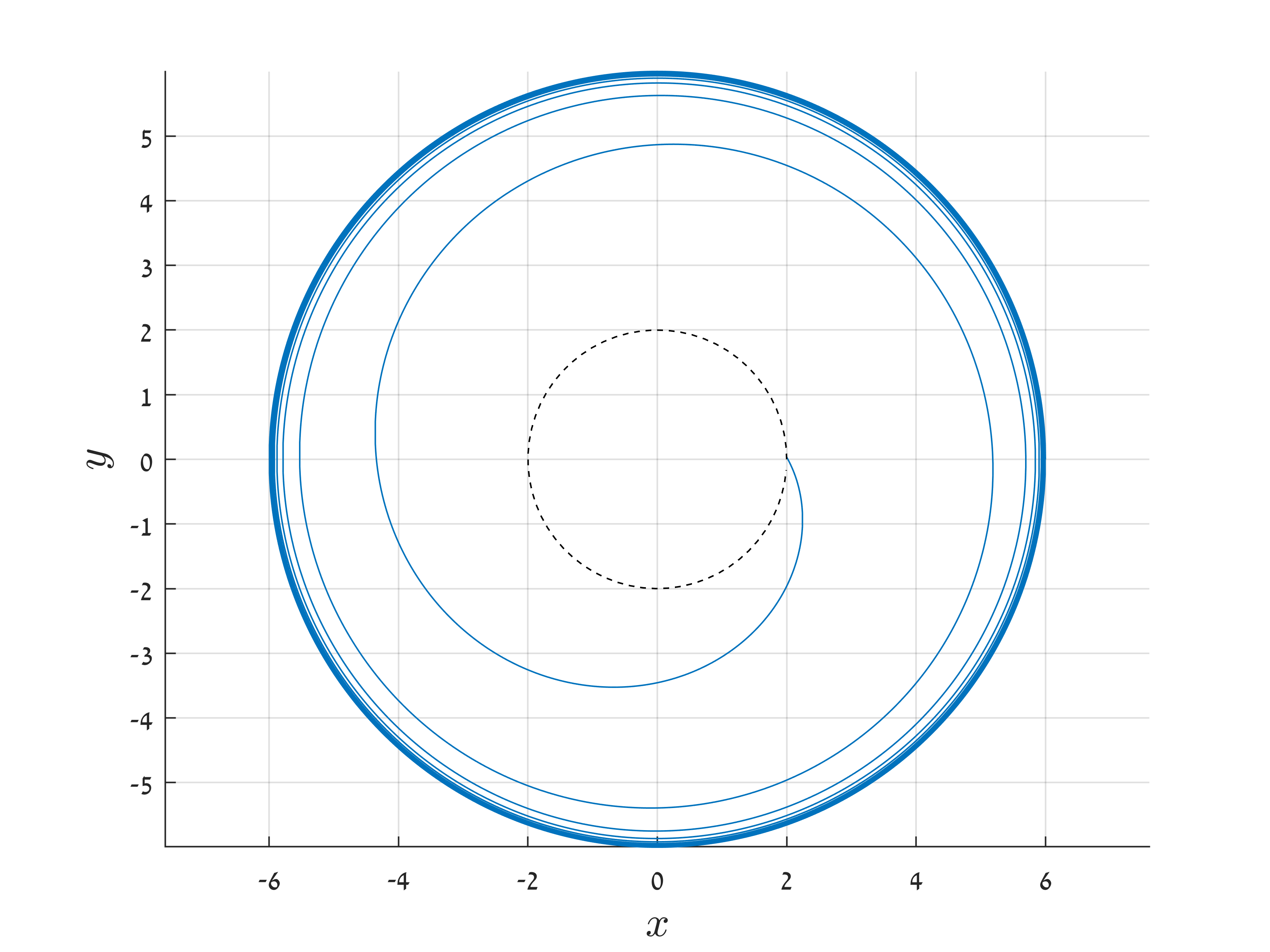}
\caption{The geodesic universal infall (GUI) trajectory of a test particle from the ISCO, $R=6M$, to the BH's horizon (black dashed line).}
\label{fig:fb1}
\end{figure}

We examine the geodesic infall of a test particle that is initially at the ISCO. Since it is a marginally stable orbit, the test particle will fall, after an infinitely long time, to the BH; The path toward the merger is schematically composed of infinite quasicircular orbits in the vicinity of the ISCO, followed by a rapid fall, during which the test particle passes most of the radial distance in $O(1)$ cycles, as demonstrated in Fig. \ref{fig:fb1}.

This orbit is a geodesic, and it is quasiuniversal, as toward the merger, small mass-ratio binary systems, which had evolved along quasicircular orbits, tend to this trajectory. Therefore, we identify it as the geodesic universal infall (GUI) trajectory.

Given the values of energy and angular momentum at the ISCO, $E_{ISCO}=\sqrt{\frac{8}{9}}$, $L_{ISCO}=\sqrt{12}M$, the GUI trajectory can be analytically calculated, yielding the implicit relation $t=g(R_0)-g(R)$, with $R_0=R(t=0)$ and
\begin{equation} \label{eq:b5}
\begin{aligned}
g(R)= & \sqrt{\frac{8R}{6-R}}\left(24-R\right)+44\sqrt{2}\sin^{-1}\left(\sqrt{\frac{6-R}{6}}\right)\\
& -4\tanh^{-1}\left(\sqrt{\frac{6-R}{2R}}\right).
\end{aligned}
\end{equation}
%
This result is equivalent to a previous result of \cite{BK11}.
\subsection{Regge-Wheeler-Zerilli equation}
The Regge-Wheeler-Zerilli (RWZ) equation, as presented below, is a wave equation with a potential, induced by the spacetime curvature, and a source term, derived from the stress-energy tensor,
\begin{equation} \label{eq:a1}
\partial^2_t \psi^{(\lambda)}_{\li,m}-\partial^2_{r^*} \psi^{(\lambda)}_{\li,m}+\mathit{V}^{(\lambda)}_\li\psi^{(\lambda)}_{\li,m}=\mathit{S}^{(\lambda)}_{\li,m},
\end{equation}
where $(\li,m)$ are the multipolar indices, $\lambda\in\{e,o\}$ is the parity, $r^*=r+2M\log\left(\frac{r}{2M}-1\right)$ is the tortoise coordinate, $\mathit{V}^{(\lambda)}_\li$ is the curvature potential and $\mathit{S}^{(\lambda)}_{\li m}$ is the source term. For the explicit expressions see Appendix A. Note that the RWZ function, $\psi^{(\lambda)}_{\li,m}$, is proportional to the asymptotic GW amplitude, where the $1/r$ dependence is factored out \cite{N05}.

\subsubsection{Energy $\&$ linear momentum fluxes}
The physical quantities associated with the GW can be determined from the RWZ function, $\psi^{(\lambda)}_{\li,m}$. Specifically, the energy and linear momentum fluxes are given by \cite{N05,M2}
\begin{equation} \label{eq:c1}
     \dot{E}=\frac{1}{8\pi}\sum_{\substack{\li\geq2\\0\leq m \leq\li}}\delta_m\frac{(\li+2)!}{(\li-2)!}\left|\dot{\psi}^{(\lambda)}_{\li,m}\right|^2,
\end{equation}
where we sum only over $m\geq0$, using the symmetry $\psi_{\li, -m}^{(\lambda)}=(-1)^m\psi_{\li,m}^{(\lambda)*}$ and denoting $\delta_m=\left\{\def\arraystretch{0.5}\begin{tabular}{@{}l@{\quad}l@{}}
  $1/2$ & $m=0$ \\
  $1$ & otherwise
\end{tabular}\right.$.
\begin{equation} \label{eq:c3}
\begin{aligned}
\dot{P}_x+i\dot{P}_y & = \frac{1}{8\pi}\sum_{\substack{\li\geq2\\0\leq m \leq\li}}\delta_m 
\left[ ia_{\li,m}\dot{\psi}^{(e)}_{\li,m}\dot{\psi}^{(o)*}_{\li,m+1}+b_{\li,m}\dot{\psi}^{(\lambda)}_{\li,m}\dot{\psi}^{(\lambda)*}_{\li+1,m+1}\right.\\
& \kern2em \left.-\left( ia_{\li,-m}\dot{\psi}^{(e)*}_{\li,m}\dot{\psi}^{(o)}_{\li,m-1}+b_{\li, -m}\dot{\psi}^{(\lambda)*}_{\li,m}\dot{\psi}^{(\lambda)}_{\li+1,m-1}\right)\right]
\end{aligned}
\end{equation}
where
$$a_{\li,m}=2(\li-1)(\li+2)\sqrt{(\li-m)(\li+m+1)}$$
$$b_{\li,m}=\frac{(\li+3)!}{(\li+1)(\li-2)!}\sqrt{\frac{(\li+m+1)(\li+m+2)}{(2\li+1)(2\li+3)}}$$
\section{Numerical Method}
Following \cite{N07}, we numerically solve the RWZ equation using a second-order Lax-Wendorff scheme, with Sommerfeld absorbing boundary conditions \cite{Numrec,r11}:
\begin{equation}
\lim_{r^*\to\pm\infty} \left(\partial_t \psi^{(\lambda)}_{\li,m}\pm\partial_{r^*} \psi^{(\lambda)}_{\li,m}\right)=0.
\end{equation}
For the initial conditions, adopting  \cite{N07} pragmatic approach, we set $\psi^{(\lambda)}_{\li,m}\left(r^*,t=0\right)=\dot{\psi}^{(\lambda)}_{\li,m}\left(r^*,t=0\right)=0$. This convenient choice leads to an unphysical initial burst that propagates outward, as can be seen in Fig. \ref{fig:fe1}. In our numerical scheme, we model the delta function in the source term as a narrow Gaussian, with a standard deviation of four grid cells. 
\subsubsection{Extrapolation to $\mathcal{I}^+$}
A conceptual limitation of the numerical calculation stems from the extraction of the GW at a finite distance. This obstacle is commonly overcome by extracting the GW at several different radii and extrapolating to null infinity, $\mathcal{I}^+$, by expanding it as a series in $1/r$ \cite{ext1,ext2,SXS}. There are other approaches for extracting the signal directly at $\mathcal{I}^+$, for example, compactification of the spatial domain \cite{N11} or characteristic extraction \cite{Ni1}. We present a different method of frequency-domain extrapolation. 

Given a time-domain signal, $\psi_{R^*}$, numerically extracted at some finite distance $R^*\gg M$, its propagation to $\mathcal{I}^+$ can be analytically calculated. At this limit, the curvature potential is approximately $V(r^*)\sim\frac{\Lambda}{r^{*2}}$, and the RWZ equation can be solved in the frequency domain\footnote{For a given ($\li,m$) and parity $\lambda$, which are omitted from the derivation to simplify the notation.\label{fn:1}}:
\begin{subequations}
\begin{gather}
\frac{\partial^2}{\partial\xi^2}\widetilde{\psi}+\left(1-\frac{\Lambda}{\xi^2}\right)\widetilde{\psi}=0 \label{eq:ext1}\\
\widetilde{\psi}=A_\omega
e^{-i\xi}\sqrt{\xi}JiY\big(\li+\frac{1}{2};\xi\big)\label{eq:ext2} 
\end{gather}
\end{subequations} 
where $\widetilde{\psi}(\omega,r^*)$ is the Fourier transform of $\psi(t,r^*)$, $\xi=\omega r^*$, $JiY(\nu;\xi)\equiv J_\nu(\xi)+iY_\nu(\xi)$ is a combination of the Bessel functions of the first and second kind, and $A_\omega$ is a $\omega$-dependant coefficient. 

The GW at $\mathcal{I}^+$ can be evaluated by taking the limit $\xi\rg\infty$:
\begin{equation}\label{eq:ext3}
\widetilde{\psi}_\infty=\widetilde{\psi}_{R^*}/\widetilde{\chi}(\xi_{R^*})
\end{equation}
where $\xi_{R^*}=\omega R^*$, $\widetilde{\psi}_{R^*}$ is the Fourier transform of the numerically extracted $\psi_{R^*}$, and $\widetilde{\chi}(\xi_R)$ is a correction function, defined as
\begin{equation}\label{eq:ext4}
\widetilde{\chi}(\xi_{R^*})=\sqrt{\frac{\pi}{2}}e^{-i\left[\xi_{R^*}-\frac{\pi}{2}\left(\li+1\right)\right]}\sqrt{\xi_{R^*}}JiY\big(\li+\frac{1}{2};\xi_{R^*}\big)
\end{equation}

\begin{figure}[ht]
\centering
  \includegraphics[width=8.6cm]{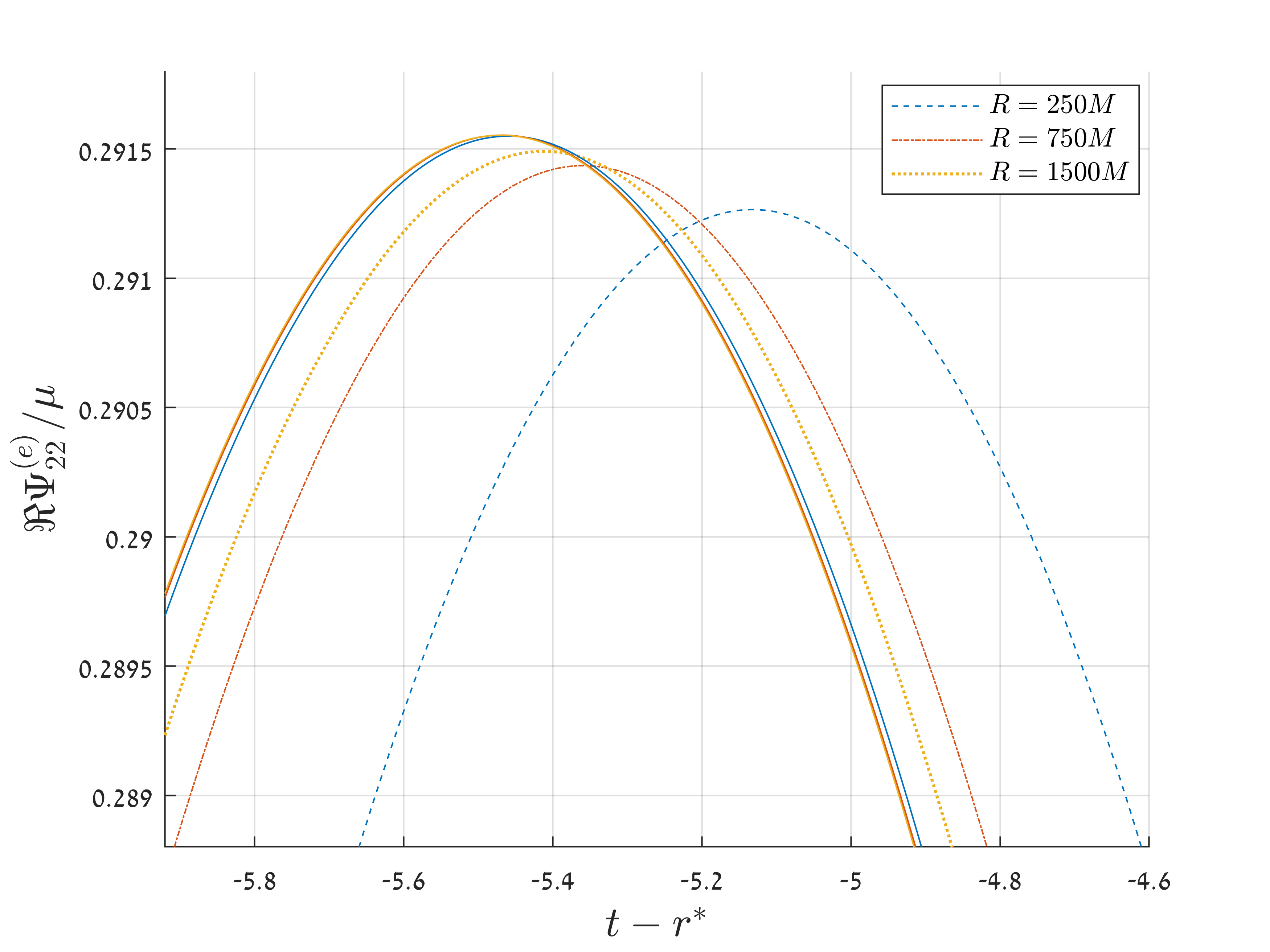}
  \caption{Comparison between the GW extracted at three different radii, $R=250M$ (blue dashed line), $R=750M$ (red dash-dotted line), and $R=1500M$ (yellow dotted line), to the GW at $\mathcal{I}^+$, independently extrapolated from each of the signals (solid lines, respective colors). The time is shifted such that each signal reaches its maximum absolute magnitude at $t-r^*=0$. We present a zoomed picture around the peak of the dominant RWZ function real part.}
\label{fig:fc2}
\end{figure}

Thus, the GW at $\mathcal{I}^+$ can be determined based on a numerical calculation of the RWZ function at a single finite distance. We examine this method by extracting the RWZ function at different radii, $R=250M$, $750M$, $1500M$, and independently extrapolating them to $\mathcal{I}^+$. As can be seen in Fig. \ref{fig:fc2}, there is a clear mismatch in amplitude and phase between the unextrapolated signals (dashed lines), which improves significantly after the extrapolation (solid lines). Quantitatively, the relative differences in the maximum absolute magnitude and phase are: $\delta A/A\sim 2.5\times10^{-3}$, $\delta\phi\sim0.069$ between $R=1500M$ and $R=250M$, which after extrapolation improves by more than an order of magnitude, $\delta A/A|_{\mathcal{I}^+}\sim 7\times10^{-5}$, $\delta\phi|_{\mathcal{I}^+}\sim0.003$, and between $R=1500M$ and $R=750M$, $\delta A/A\sim 5\times10^{-4}$, $\delta\phi\sim0.067$, and after extrapolation $\delta A/A|_{\mathcal{I}^+}\sim 6\times10^{-6}$, $\delta\phi|_{\mathcal{I}^+}\sim5\times10^{-4}$. We see the same consistent alignment of the extrapolated signals in higher multipoles as well.
\section{Results}
We implement our numerical method to calculate the GW and emitted fluxes from a source that moves along a geodesic in the Schwarzschild geometry. We begin by reconstructing known results for circular orbits, as a validation test, and then continue to estimate the emission from a source that moves along the GUI trajectory.
\subsection{Circular orbits}
We test the numerical calculation scheme in the well-studied circular orbit case \cite{M03,SL06,N10,N11}. For comparison with the literature, we examine the GW emission from a circular orbit at distance $R=7.9456M$. We calculate $\psi^{(\lambda)}_{\li,m}$ up to $\li=8$, and determine the corresponding energy flux, using Eq. (\ref{eq:c1}). In addition, we develop a semianalytical method to calculate the GW from circular orbits, as discussed in the following section. We get very good agreement between the results of the numerical calculation, the semi-analytical method, and known results in the literature \cite{N11}, as summarized in Table \ref{table:d1} of Appendix B.
\subsubsection{Comparison to semianalytical solution}
The RWZ source term has in the circular orbit case a simple time dependence, $\mathit{S}\propto e^{-im\Omega t}$, where $\Omega=\sqrt{M/R^{3}}$ is the orbital frequency. Substituting the ansatz $\psi(r,t)=f(r)e^{-im\Omega t}$, gives an homogeneous ODE for $f(r)$\footnotemark[\value{footnote}]:
\begin{equation} \label{eq:d7}
\begin{aligned}
     \frac{d^2f}{dr^{*2}}+\Big( m^2\Omega^2-\mathit{V}\Big) f=0.
\end{aligned}
\end{equation}

\begin{figure}[h!]
\centering
\includegraphics[width=8.6cm]{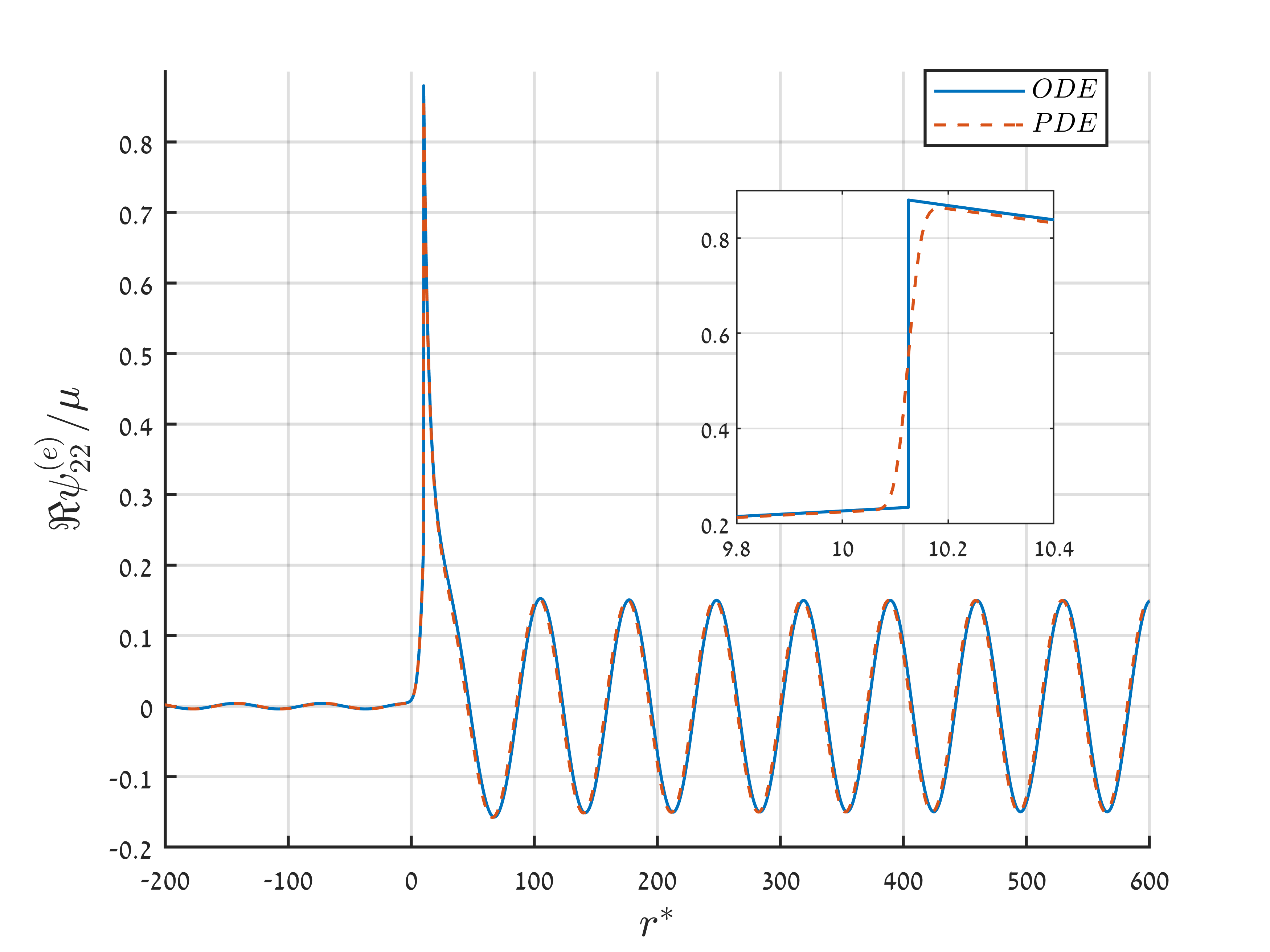}
\caption{Comparison between the numerical PDE solution of Eq. (\ref{eq:a1}), and the semianalytical ODE solution of Eq. (\ref{eq:d7}), for a circular orbit at $R=7.9456M$. We present the dominant RWZ function. The small frame displays an enlarged picture around the discontinuity at $R^*$, where the two solutions slightly deviate due to the smoothing of the delta function in the PDE numerical scheme.}
\label{fig:fd2}
\end{figure}

Thus, the original problem reduces to solving Eq. (\ref{eq:d7}) in two separate regimes, $r^*<R^*$ and $r^*>R^*$. A unique solution is obtained by imposing outgoing wave boundary conditions and specific finite discontinuity conditions at $r^*=R^*$: $\left\{\def\arraystretch{1}\begin{tabular}{@{}l}
  $\Delta f|_{R^*} =\widetilde{D}(R)F(R)$ \\
  $\Delta f'|_{R^*}=\widetilde{D}(R)G(R)$
\end{tabular}\right.$, where $\widetilde{D}$, $F$, and $G$ are given in Appendix A. The ODE solution is in a good agreement with the full numerical one, as can be seen in Fig. \ref{fig:fd2}.

Asymptotically, for $r\rg\infty$, $\psi\rg\beta e^{-im\Omega (t-r^*)}$ , where $\beta$ is a complex coefficient determined by the discontinuity conditions. By substituting this expression into Eqs. (\ref{eq:c1}) and (\ref{eq:c3}), we get the energy and the linear momentum fluxes\footnote{The angular momentum flux can be calculated in a similar manner and the known relation $\frac{dE}{dt}=\Omega\frac{dJ}{dt}$ can be easily derived.}, as a functions of $\beta$. Thus, for example, the energy flux can be written as
\begin{equation} \label{eq:d11}
\dot{E}_{\li,m}=\frac{1}{8\pi}\frac{(\li+2)!}{(\li-2)!}\frac{m^2}{R^3}\left|\beta_{\li,m}\right|^2
\end{equation}

In Appendix B, we provide an analytical solution for $\beta$ at the Newtonian limit, $R\gg M$, which allows for reconstructing the known quadrupole radiation formula \cite{LL}, $\dot{E}_{22} =\frac{32}{5}\frac{M^3\mu^2}{R^5}$. Moreover, we show that asymptotically, the contribution of the high multipoles to the emitted fluxes decays exponentially at a constant, radius-dependent rate:
\begin{equation}
\frac{\dot{E}_{\li+1}}{\dot{E}_{\li}}\ \sim \ \frac{\dot{P}_{\li+1}}{\dot{P}_{\li}}\sim \frac{e^2}{4R}
\label{eq:cp}
\end{equation}
for $\ell \gg 1$.
\subsection{Inspiral $\&$ merger}
\begin{figure}[ht]
\centering
  \includegraphics[width=8.6cm]{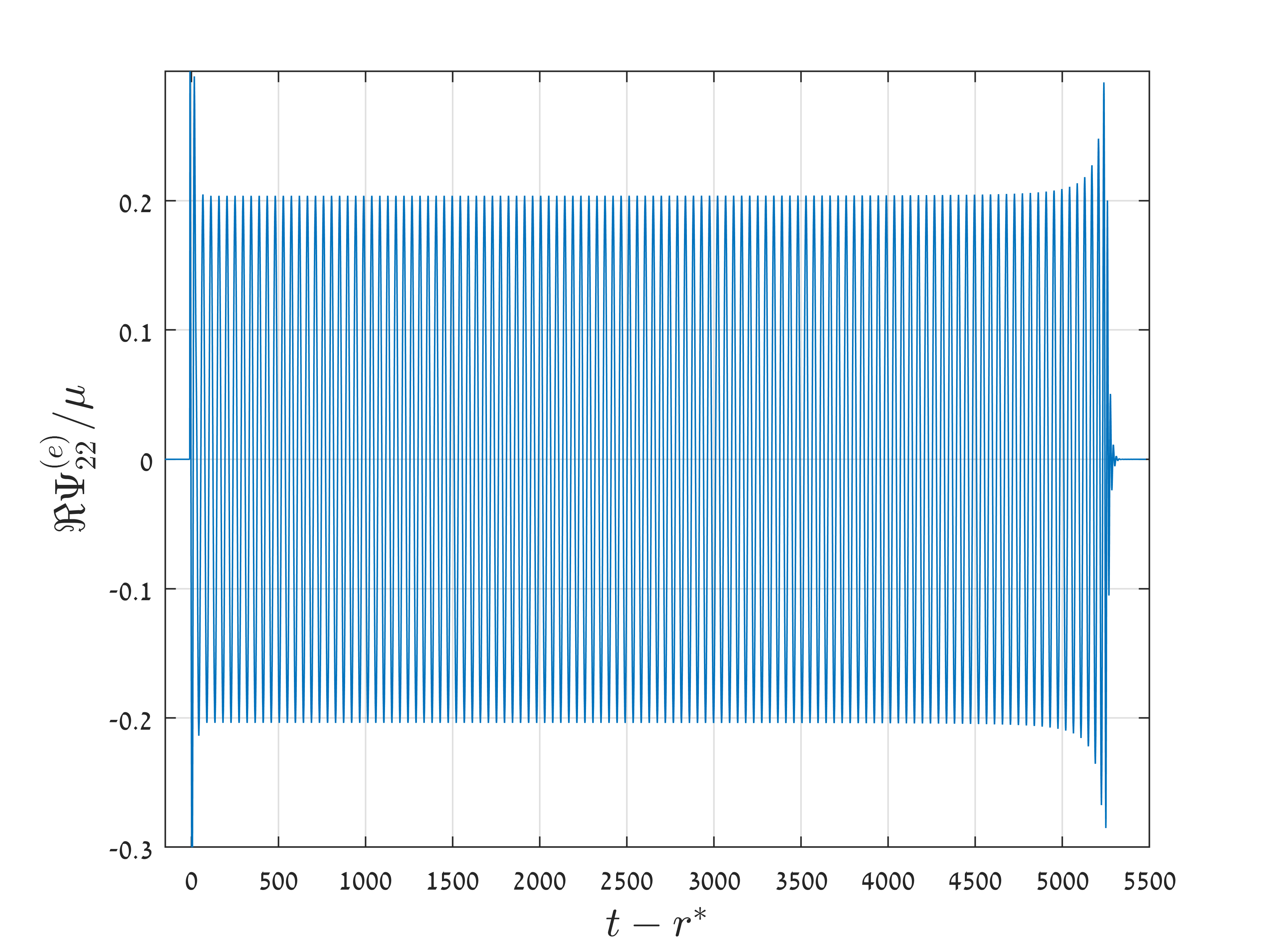}
  \caption{The GUI waveform: binary merger at the test-particle limit: the real part of the dominant RWZ function, extrapolated to $\mathcal{I}^+$, as a function of the retarded time. After the initial induced burst, we see the typical binary merger "chirp" signal.}
\label{fig:fe1}
\end{figure}

We now move to calculate the merger waveform by numerically solving the RWZ equation for a test particle that moves along the GUI trajectory. A similar methodology of calculating the GW emission from a source that moves along a geodesic has been adopted in \cite{BK11a, fr1}. Qualitatively, after the initial induced burst, the signal oscillates, with approximately constant amplitude and frequency, corresponding to the quasicircular orbits in the vicinity of the ISCO. Then, it sharply increases in amplitude and frequency due to the particle's rapid infall at the last few orbits, and quickly decays. Thus, for example, Fig. \ref{fig:fe1} presents the dominant, quadrupole RWZ function. As can be seen, the amplitude increases by about $45\%$ at its peak, compared to its initial magnitude during the quasicircular orbits.
\subsubsection{The GUI waveform}
The universality of the GUI trajectory entails that the test-particle GUI waveform, as presented in Fig. \ref{fig:fe1}, can be used as a general template for small mass-ratio binary mergers. The GUI waveform captures the GW emission pattern from the merger back to the ISCO crossing since the secondary BH's deviation from the GUI trajectory, due to the GW emission, does not yield a significant phase difference between the GUI waveform and the small mass-ratio waveform during the infall from the ISCO to the horizon.

The two signals, the GUI and the small mass-ratio one, go out of phase on a timescale that corresponds to the orbital frequency change during the GW-driven inspiral: $\ddot{\Phi}t^2\sim\dot{R}t^2\sim1\rg t\propto\nu^{-3/10}$, where we used that the radial velocity at the vicinity of the ISCO scales as $\dot{R}\propto\nu^{3/5}$ \cite{D00,OT00}. Thus, we can point out three relevant timescales: the inspiral timescale, $t\propto\nu^{-1}$, the above mentioned dephasing timescale, $t\propto\nu^{-3/10}$, and the plunge timescale, $t\propto\nu^{-1/5}$ \cite{D00,OT00}.

\begin{figure}[ht]
\centering
  \includegraphics[width=8.6cm]{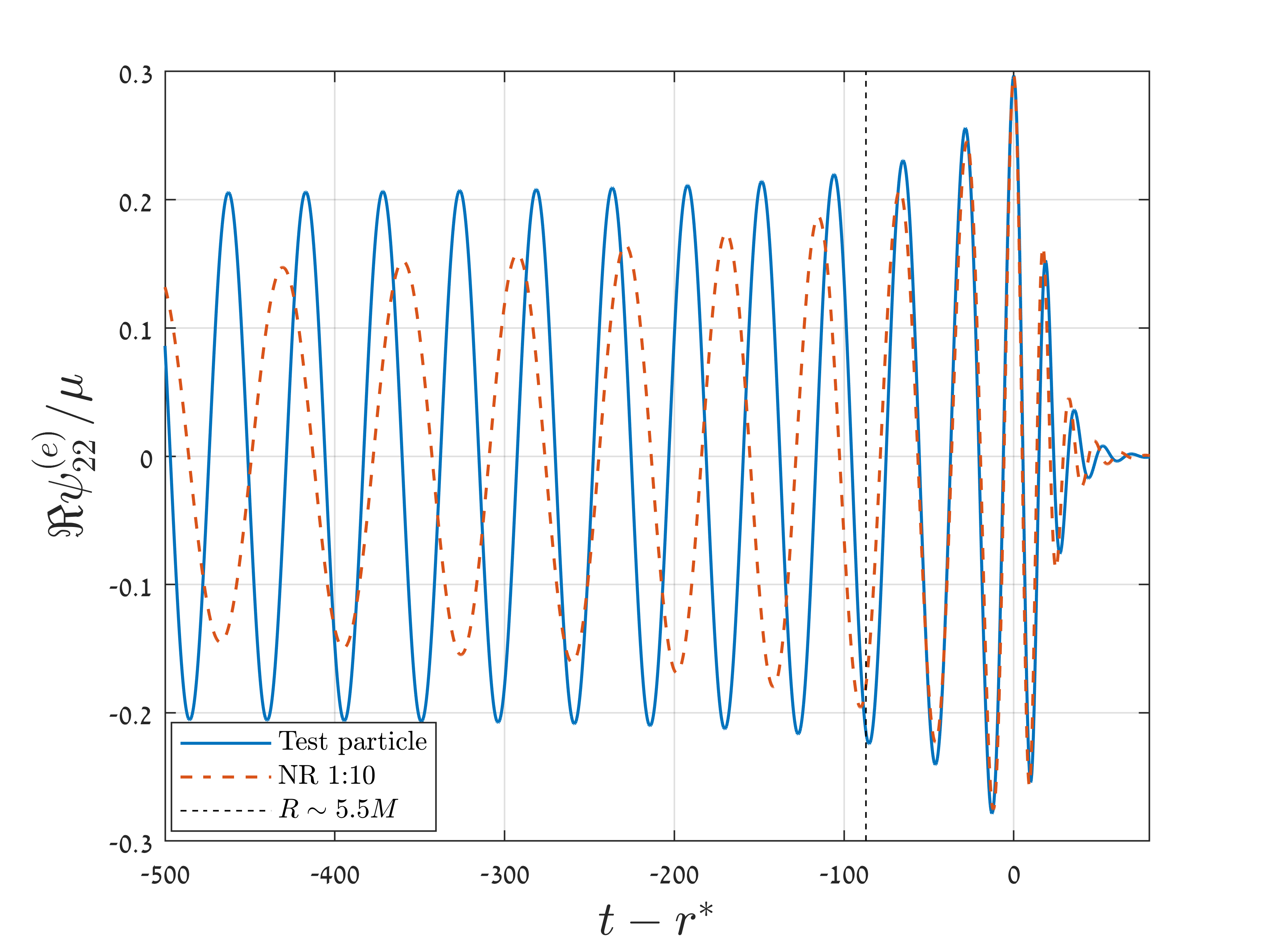}
  \caption{Comparison between the GUI waveform (blue line) and NR simulation waveform, from the SXS catalog \cite{SXS10}, with mass ratio $1:10$ (red dashed line). Both of the signals are shifted so that their real part reaches its maximum value at $t-r^*=0$. The vertical dashed line corresponds to the time when the test particle is at $R\sim5.5M$ and the relative distance in the NR simulation is about $6.15M$.}
\label{fig:fe2}
\end{figure}

As a preliminary proof of concept, we compare the GUI waveform to a waveform from a NR simulation of a binary merger with mass ratio 1:10 \cite{SXS10}, as presented in Fig. \ref{fig:fe2}. We get a remarkably good agreement between the maximum amplitudes of the two signals, with a relative difference of $0.3\%$, and they stay in phase for about three cycles.
\subsubsection{Linear momentum - recoil Velocity}
Using Eq. (\ref{eq:c3}), we calculate the emitted linear momentum flux. The recoil velocity is then obtained by integration:
\begin{equation} \label{eq:e1} V_x+iV_y=V_0-\frac{1}{M}\int^t_{-\infty} \left(\dot{P_x}+i\dot{P_y}\right)dt'
\end{equation}

First, we point out the scaling. As can be seen in Appendix A, the source term in the RWZ equation is linear in $\mu$. Therefore, $\psi\propto\mu$ and its time derivative scales as $\dot{\psi}\propto\frac{\mu}{M}\equiv\nu$. Given this scaling, Eqs. (\ref{eq:c1}) and (\ref{eq:c3}) imply that the radiated fluxes scale as $\nu^2$, and so does the recoil velocity.

The integration constant, $V_0$, is determined by the requirement that the initially oscillating velocity, corresponding to the GW emission along the quasicircular orbits, will have zero mean, as discussed in \cite{N10}. For comparison with the literature, we note that up to $\li=7$ the recoil magnitude is $V^{(7)}/\nu^2=0.0455$, which is about $1-2\%$ larger than the results of \cite{N10,N11}, which were calculated for a small, finite mass ratios. Our results are in a closer agreement to that of \cite{H10,N14}. Further analysis needs to be done to establish if this slight deviation stems from numerical inaccuracies or from the fact that our result is calculated along the exact geodesic trajectory.

The higher multipoles have a decreasing, yet significant, contribution to the total recoil velocity \cite{M03,H06,H10,N10}, as can be seen in Fig. \ref{fig:f_v2} and in Appendix C, where we present the detailed results for each multipole separately. Therefore, we wish to evaluate the contribution of the infinite ``tail" of high multipoles,
\begin{equation} \label{eq:e2}
V=V^{(L)} + \sum_{\li=L+1}^{\infty}{\delta V^{(\li)}},
\end{equation}
where $L$ is the highest multipole that was numerically calculated; for this work, $L=10$. Based on the results for circular orbits as a heuristic guideline, and reinforced by the results of \cite{D14} regarding the QNM energy flux exponential decay, we assume that the contribution of the higher multipoles along the merger scenario decreases exponentially as well:
\begin{equation} \label{eq:e4}
\delta V^{(l)}=a\times C^\li.
\end{equation}
\begin{figure}[ht] 
\centering
\includegraphics[width=8.6cm]{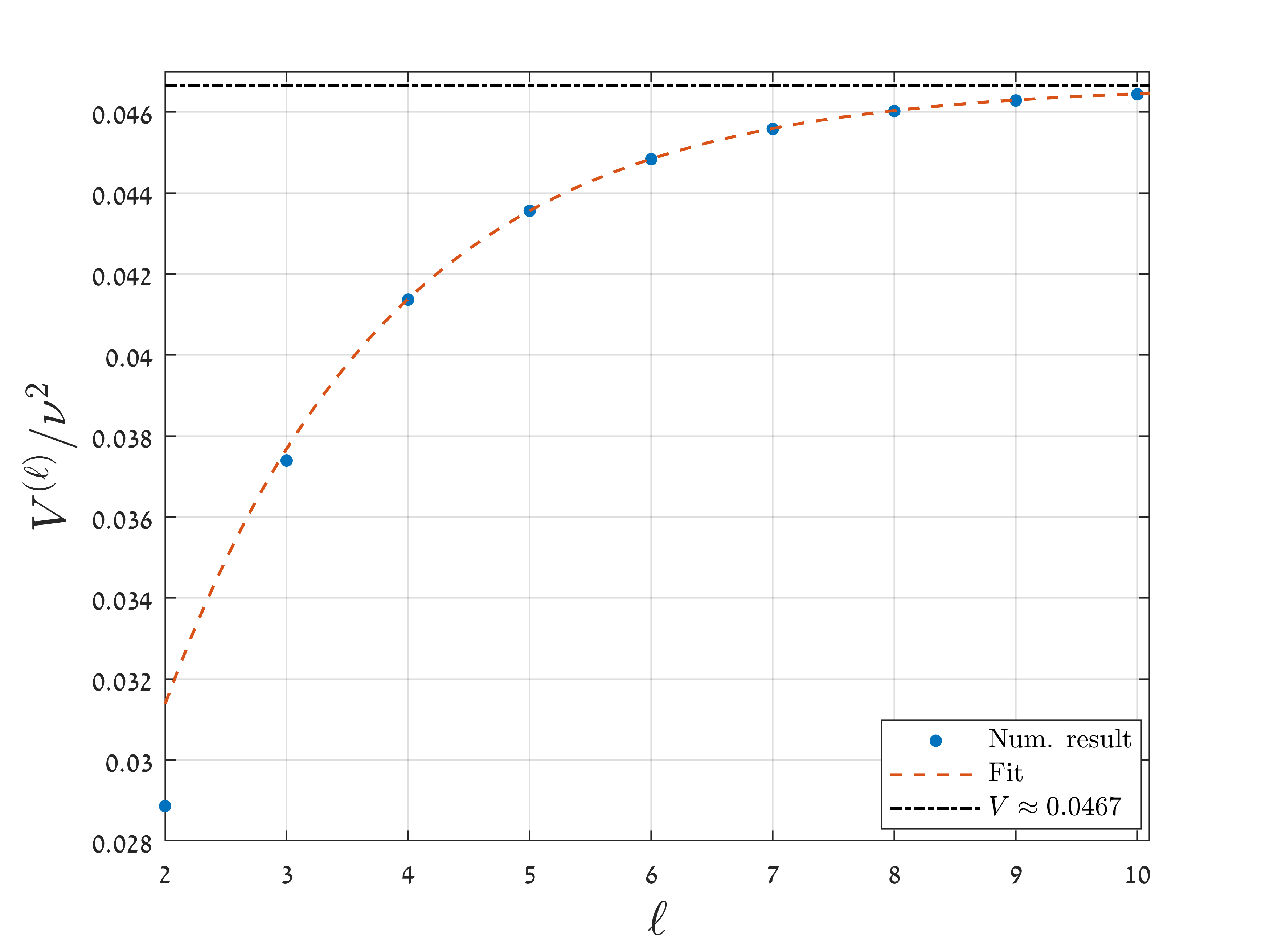}
\caption{The final recoil velocity, up to a given multipole $\li$. The total velocity converges to $V/\nu^2\approx0.0467$ (black dash-dotted line) and corresponds to an exponential decay trend (red dashed line). The detailed numerical results are presented in Table \ref{table:f1} of Appendix B.}
\label{fig:f_v2}
\end{figure}

\begin{figure}[ht] 
\centering
\includegraphics[width=8.6cm]{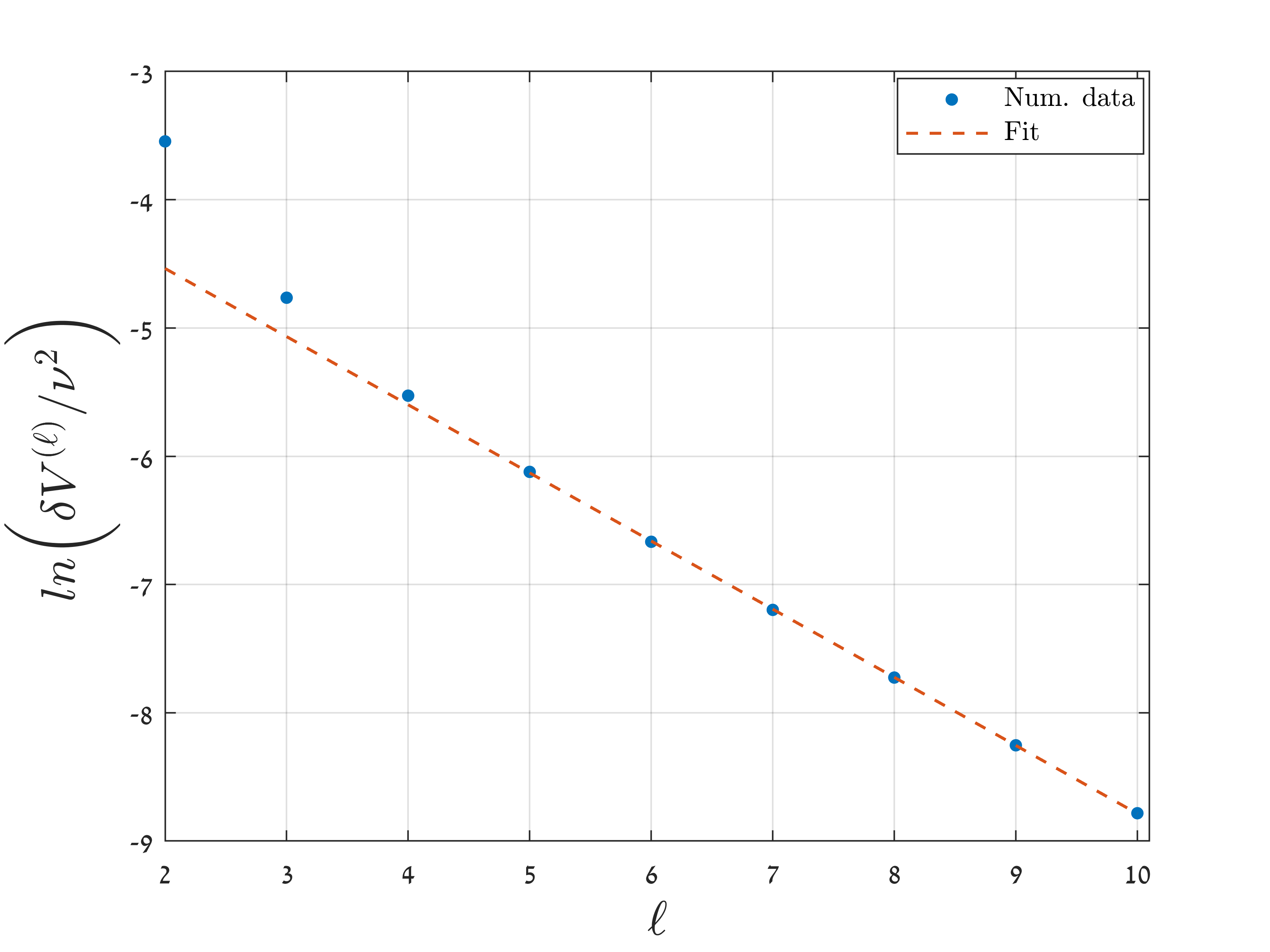}
\caption{An exponential decay fit in a semilogarithmic scale to the relative contribution of each multipole to the total recoil velocity.}
\label{fig:f_log}
\end{figure}

We get a good correspondence between the numerical results to an exponential decay trend, as can be seen in Figs. \ref{fig:f_v2} and \ref{fig:f_log}. Using the numerical fit, we determine the values of the coefficients in Eq. (\ref{eq:e4}) and evaluate the total recoil velocity: $V/\nu^2= 0.0467$. This value is about $4\%$ larger than previous results in the literature, which estimated $V/\nu^2\approx0.044$. This difference originates mostly due to the summation by extrapolation of the high multipoles contribution.
\subsubsection{Energy and Angular momentum}
We present a brief derivation of the postmerger energy and angular momentum of the remnant BH in leading order of the mass ratio. Up to the ISCO, the emitted energy and angular momentum can be determined directly: $\Delta E/\nu=\left(1-E_{ISCO}\right)=\left(1-\sqrt{\frac{8}{9}}\right)$, $\Delta J/\nu=-J_{ISCO}M=-\sqrt{12}M^2$.

The time from the ISCO crossing to the merger scales as $t \propto \nu^{-1/5}$ \cite{D00,OT00}, and as mentioned above, $\dot{E}\propto \nu^2$. Therefore, the total emitted energy after the ISCO crossing scales as $\nu^{9/5}$, and so does the emitted angular momentum. Hence, the GW emission after the ISCO crossing contributes only to the next order in the mass ratio. In summary, we get that the final energy and spin of the remnant BH are
\begin{subequations}
\begin{gather}
M_f/M  =1-\left(1-\sqrt{\frac{8}{9}}\right)\nu+O(\nu^{9/5})\label{eq:ef1}\\
a\equiv J/M^2 =\sqrt{12}\nu+O(\nu^{9/5}).\label{eq:ef2}
\end{gather}
\end{subequations}

The first order terms are in accordance with known results in the literature \cite{D14,L10}. However, the scaling of the next order, $O(\nu^{9/5})$, which was already derived by \cite{D00,OT00}, differs from the broadly used $O(\nu^2)$.
\section{Conclusions}
We present a thorough investigation of the leading order effects of a binary BH merger in the extreme mass ratio limit. We develop a new approach that allows us to perform the calculation directly in the test-particle limit, without introducing any finite mass-ratio. This can be done due to the universal characteristics of the plunge, specifically its tendency toward the GUI trajectory. In addition, this straightforward approach allows us to construct a universal waveform that describes well the peak GW emission at the final stages of small mass-ratio binary mergers. This GUI waveform may be used as a computationally inexpensive template in the ongoing search for GW from intermediate mass-ratio binary systems in earth-based detectors.

At last, using this formalism, we show that the recoil velocity has a quadratic dependence on the mass ratio and is given by $V/c\approx 0.0467\nu^2$. This result is larger than the known value in the literature by about $4\%$, mostly as a result of the high multipole contributions.
As for the final energy and spin of the remnant BH, we derive analytically the first order, linear in the mass ratio, terms and point out the scaling of the next order, $O(\nu^{9/5})$, which differs from the broadly used value in the literature. Our calculation can be generalized to spinning BH.

\begin{acknowledgments}
The authors would like to thank Scott Hughes, Amos Ori, and Tsvi Piran for useful comments. This research was partially supported by an ISF grant.
\end{acknowledgments}

\appendix
\section{SOURCE TERM AND CURVATURE POTENTIAL IN THE RWZ EQUATION}
The explicit form of the curvature potential and the source term, for a test particle, in the RWZ equation are as follows \cite{N05}:
\begin{equation} \label{eq:ap2}
\begin{aligned}
\mathit{S}^{(\lambda)}= D^{(\lambda)}&\big(R,\Theta,\Phi\big)\Big[G^{(\lambda)}\big(R\big)\delta\big( r^*-R^*\big)\\
&+F^{(\lambda)}\big(R\big)\partial_{r^*}\delta\big( r^*-R^*\big)\Big],   
\end{aligned}
\end{equation}
where $\Big(R(t),\Theta(t),\Phi(t)\Big)$ are the test particle's coordinates. Without loss of generality, we can assume that the motion is equatorial and so $\Theta(t)=\frac{\pi}{2}$. In the following equations, $r$, $t$, and $L$ are measured in units of the primary BH mass, $M$.
\subsection{Even parity perturbations}
\begin{equation} \label{eq:ap1}
\centering
\mathit{V}^{(e)}_\li=\lp \A\rp \frac{(\Lambda-2)^2(\Lambda r+6)r^2+36(\Lambda-2)r+72}{M^2r^3[(\Lambda-2)r+6]^2},  
\end{equation}
where $\Lambda=\li(\li+1)$.
\begin{equation} \label{eq:ap3}
D^{(e)}=\mu\left(\AR\right)\frac{8\pi Y_{\li m}^* \big(\Theta,\Phi\big)}{E\Lambda R \big[\left(\Lambda-2\right)R+6\big]}
\end{equation}
\begin{equation} \label{eq:ap4}
\begin{aligned}
G^{(e)}=& \frac{1}{M\left[\left(\Lambda-2\right)R+6\right]}\Bigg( -12+8\left(1-3E^2+\Lambda\right) R\\
&\ +\left(\Lambda-2\right)\Lambda R^2 -4imL \big[\left(\Lambda-2\right) R+6\big]V_\tau \Bigg. \\
&\  \Bigg. +\frac{2L^2}{R^2\left(\Lambda-2\right)}\Bigg\{\Lambda^3R^2-\Lambda^2R\big[-12+\left(5+m^2\right) R\big]\\
&\quad\quad -4\big[3-6\Lambda+m^2\left( R-3\right)^2-3R+R^2\big] \Bigg. \Bigg.\\
&\quad\quad \Bigg.\Bigg. -2\Lambda\big[3\left(5+2m^2\right)R-2\left(2+m^2\right) R^2\big]\Bigg\} \Bigg),
\end{aligned}
\end{equation}
where $V_\tau=-\sqrt{E^2-\left(\AR\right)\left(1+\frac{L^2}{R^2}\right)}$ is the radial velocity with respect to the particle's proper time.
\begin{equation} \label{eq:ap5}
F^{(e)}=-2\left(L^2+R^2\right).
\end{equation}
%
\subsection{Odd parity perturbations}
\begin{equation} \label{eq:ap6}
\mathit{V}^{(o)}_\li=\frac{1}{M^2}\lp\A\rp\lp\frac{\Lambda}{r^2}-\frac{6}{r^3}\rp
\end{equation}
\begin{equation} \label{eq:ap8}
D^{(o)}=\mu\lp\AR\rp\frac{16\pi L \partial_\theta Y_{\li m}^* \big(\Theta,\Phi\big)}{E^2R^3\Lambda(\Lambda-2)}
\end{equation}
\begin{equation} \label{eq:ap9}
\begin{aligned}
    G^{(o)} =\frac{1}{M}&\left[\frac{L^2}{R^2}\big(2R-5\big)+imLV_\tau \right.\\
    &\quad \left. +R\left(1-\frac{3}{R}-2E^2\right)\right]
\end{aligned}
\end{equation}
\begin{equation} \label{eq:ap10}
F^{(o)}=L^2+R^2.
\end{equation}
In the semianalytical method for circular orbits, as discussed in Section $IV.A.1$, we introduced $\widetilde{D}^{(\lambda)}\equiv D^{(\lambda)}e^{im\Phi}$, which factors out the $\Phi$ dependence.
\section{CIRCULAR ORBITS}
We present in Table \ref{table:d1} a comparison between the radiated energy fluxes, up to $\li=8$, as calculated using the semianalytical ODE method, the numerical PDE scheme, extrapolated to null infinity and extracted at finite radii, and known results in the literature \cite{N11}. We note that as $\li$ increases, the amplitude of the $m\ll\li$ moments becomes increasingly smaller and so more distorted by numerical noise and rescattering of the initial junk radiation. However, their contribution to the total emitted fluxes is negligible.
\begingroup
\begin{table*}
\begin{center}
\renewcommand{\arraystretch}{0.5} 
\begin{tabular}{|c | c| c | c | c | c | c |}
 \hline
 $\li$ & $m$ & $\dot{E}/\nu^2|_{ODE}$ & $\dot{E}/\nu^2|_{PDE_{\mathcal{I}^+}}$ & $\dot{E}/\nu^2|_{PDE_{1500}}$ & $\dot{E}/\nu^2|_{PDE_{250}}$ & $\dot{E}/\nu^2|_{Bernuzzi+11'}$ \\
\hline
\multirow{2}{0.5em}{$2$} & $1$ & $8.1628\times10^{-07}$	& $8.1664\times10^{-07}\ [0.04\%]$ & $8.1718\times10^{-07}\ [0.11\%]$ & $8.3622\times10^{-07}\ [2.44\%]$ & $8.1632\times10^{-07}\ [<0.01\%]$ \\
& $2$ & $1.7062\times10^{-04}$ & $1.7061\times10^{-04}\ [<0.01\%]$ & $1.7064\times10^{-04}\ [0.01\%]$ & $1.7165\times10^{-04}\ [0.60\%]$ & $1.7065\times10^{-04}\ [0.02\%]$ \\
\hline
\multirow{3}{0.5em}{$3$} & $1$ & $2.1730\times10^{-09}$ & $2.1749\times10^{-09}\ [0.09\%]$ & $2.1778\times10^{-09}\ [0.22\%]$ & $2.2833\times10^{-09}\ [5\%]$ & $2.1740\times10^{-09}\ [0.04\%]$ \\
& $2$ & $2.5198\times10^{-07}$ & $2.5196\times10^{-07}\ [0.01\%]$ & $2.5205\times10^{-07}\ [0.03\%]$ & $2.5506\times10^{-07}\ [1.22\%]$ & $2.5203\times10^{-07}\ [0.02\%]$ \\
& $3$ &	$2.5470\times10^{-05}$ & $2.5475\times10^{-05}\ [0.02\%]$ & $2.5479\times10^{-05}\	[0.03\%]$ & $2.5609\times10^{-05}\ [0.54\%]$ & $2.5481\times10^{-05}\ [0.04\%]$ \\
\hline
\multirow{4}{0.5em}{$4$} & $1$ & $8.3947\times10^{-13}$ & $8.4071\times10^{-13}\ [0.15\%]$ & $8.4256\times10^{-13}\ [0.37\%]$ & $9.1434\times10^{-13}\ [9\%]$ & $8.4001\times10^{-13}\ [0.06\%]$ \\
& $2$ &	$2.5089\times10^{-09}$ & $2.5088\times10^{-09}\ [0.01\%]$ & $2.5102\times10^{-09}\	[0.05\%]$ & $2.5608\times10^{-09}\ [2.07\%]$ & $2.5115\times10^{-09}\ [0.10\%]$ \\
& $3$ &	$5.7748\times10^{-08}$ & $5.7766\times10^{-08}\ [0.03\%]$ & $5.7780\times10^{-08}\	[0.05\%]$ & $5.8274\times10^{-08}\ [0.91\%]$ & $5.7777\times10^{-08}\ [0.05\%]$ \\
& $4$ & $4.7252\times10^{-06}$ & $4.7259\times10^{-06}\ [0.01\%]$ & $4.7265\times10^{-06}\	[0.03\%]$ & $4.7499\times10^{-06}\ [0.52\%]$ & $4.7289\times10^{-06}\ [0.08\%]$ \\
\hline
\multirow{5}{0.5em}{$5$} & $1$	& $1.2594\times10^{-15}$ & $1.2621\times10^{-15}\ [0.22\%]$ & $1.2663\times10^{-15}\ [0.55\%]$ & $1.4398\times10^{-15}\ [14\%]$ & $1.2612\times10^{-15}\ [0.14\%]$ \\
& $2$ &	$2.7895\times10^{-12}$ & $2.7891\times10^{-12}\ [0.01\%]$ & $2.7914\times10^{-12}\	[0.07\%]$ & $2.8772\times10^{-12}\ [3.14\%]$ & $2.7925\times10^{-12}\ [0.11\%]$ \\
& $3$ & $1.0932\times10^{-09}$ & $1.0938\times10^{-09}\ [0.06\%]$ & $1.0942\times10^{-09}\	[0.09\%]$ & $1.1083\times10^{-09}\ [1.39\%]$ & $1.0948\times10^{-09}\ [0.15\%]$ \\
& $4$ & $1.2324\times10^{-08}$ & $1.2325\times10^{-08}\ [0.02\%]$ & $1.2328\times10^{-08}\	[0.04\%]$ & $1.2420\times10^{-08}\ [0.78\%]$ & $1.2334\times10^{-08}\ [0.09\%]$ \\
& $5$ &	$9.4556\times10^{-07}$ & $9.4593\times10^{-07}\ [0.04\%]$ & $9.4605\times10^{-07}\	[0.05\%]$ & $9.5041\times10^{-07}\ [0.51\%]$ & $9.4660\times10^{-07}\ [0.11\%]$ \\
\hline
\multirow{6}{0.5em}{$6$} & $1$ & $2.8718\times10^{-19}$ & $2.8807\times10^{-19}\ [0.31\%]$ & $2.8941\times10^{-19}\ [0.78\%]$ & $3.5025\times10^{-19}\ [22\%]$ & $2.9141\times10^{-19}\ [1.47\%]$ \\
& $2$ & $1.3338\times10^{-14}$ & $1.3337\times10^{-14}\ [0.01\%]$ & $1.3352\times10^{-14}\	[0.11\%]$ & $1.3938\times10^{-14}\ [4.50\%]$ & $1.3368\times10^{-14}\ [0.23\%]$ \\
& $3$ &	$1.9644\times10^{-12}$ & $1.9657\times10^{-12}\ [0.07\%]$ & $1.9667\times10^{-12}\	[0.12\%]$ & $2.0027\times10^{-12}\ [1.95\%]$ & $1.9677\times10^{-12}\ [0.17\%]$ \\
& $4$ & $3.4949\times10^{-10}$ & $3.4962\times10^{-10}\ [0.04\%]$ & $3.4972\times10^{-10}\	[0.06\%]$ & $3.5337\times10^{-10}\ [1.11\%]$ & $3.5023\times10^{-10}\ [0.21\%]$ \\
& $5$ & $2.5698\times10^{-09}$ & $2.5710\times10^{-09}\ [0.05\%]$ & $2.5715\times10^{-09}\ [0.07\%]$ & $2.5881\times10^{-09}\ [0.71\%]$ & $2.5728\times10^{-09}\ [0.12\%]$ \\
& $6$ & $1.9598\times10^{-07}$ & $1.9606\times10^{-07}\ [0.04\%]$ & $1.9608\times10^{-07}\	[0.05\%]$ & $1.9698\times10^{-07}\ [0.51\%]$ & $1.9621\times10^{-07}\ [0.12\%]$ \\
\hline
\multirow{7}{0.5em}{$7$} & $1$ & $2.4477\times10^{-22}$ & $2.4578\times10^{-22}\ [0.41\%]$ & $2.4730\times10^{-22}\ [1.03\%]$ & $3.2534\times10^{-22}\ [33\%]$ & $\dots$ \\
& $2$ & $9.2470\times10^{-18}$ & $9.2450\times10^{-18}\ [0.02\%]$ & $9.2592\times10^{-18}\	[0.13\%]$ & $9.8131\times10^{-18}\ [6.12\%]$ & $9.2734\times10^{-18}\ [0.29\%]$ \\
& $3$ & $1.7391\times10^{-14}$ & $1.7410\times10^{-14}\ [0.11\%]$ & $1.7422\times10^{-14}\	[0.18\%]$ & $1.7850\times10^{-14}\ [2.64\%]$ & $1.7446\times10^{-14}\ [0.32\%]$ \\
& $4$ & $8.1842\times10^{-13}$ & $8.1872\times10^{-13}\ [0.04\%]$ & $8.1903\times10^{-13}\	[0.08\%]$ & $8.3052\times10^{-13}\ [1.48\%]$ & $8.2034\times10^{-13}\ [0.23\%]$ \\
& $5$ & $9.7236\times10^{-11}$ & $9.7312\times10^{-11}\ [0.08\%]$ & $9.7336\times10^{-11}\	[0.10\%]$ & $9.8178\times10^{-11}\ [0.97\%]$ & $9.7500\times10^{-11}\	[0.27\%]$ \\
& $6$ &	$5.3160\times10^{-10}$ & $5.3184\times10^{-10}\ [0.04\%]$ & $	5.3193\times10^{-10}\ [0.06\%]$ & $5.3520\times10^{-10}\ [0.68\%]$ & $5.3226\times10^{-10}\ [0.12\%]$ \\
& $7$ &	$4.1390\times10^{-08}$ & $4.1416\times10^{-08}\ [0.06\%]$ & $4.1421\times10^{-08}\	[0.08\%]$ & $4.1603\times10^{-08}\ [0.52\%]$ & $4.1414\times10^{-08}\ [0.06\%]$ \\
\hline
\multirow{8}{0.5em}{$8$} & $1$ & $3.5076\times10^{-26}$ & $8.9236\times10^{-26}\ [150\%]$ & $9.3647\times10^{-26}\ [160\%]$ & $1.0378\times10^{-25}\ [200\%]$ & $\dots$ \\
& $2$ & $2.5726\times10^{-20}$ & $2.5724\times10^{-20}\ [0.01\%]$ & $2.5775\times10^{-20}\	[0.19\%]$ & $2.7812\times10^{-20}\ [8.11\%]$ & $2.8445\times10^{-20}\ [10.57\%]$ \\
& $3$ & $2.0951\times10^{-17}$ & $2.0976\times10^{-17}\ [0.12\%]$ & $2.0995\times10^{-17}\ [0.21\%]$ & $2.1667\times10^{-17}\ [3.42\%]$ & $2.1027\times10^{-17}\ [0.36\%]$ \\
& $4$ &	$1.0870\times10^{-14}$ & $1.0877\times10^{-14}\ [0.06\%]$ & $1.0882\times10^{-14}\	[0.11\%]$ & $1.1080\times10^{-14}\ [1.93\%]$ & $1.0914\times10^{-14}\ [0.40\%]$ \\
& $5$ & $2.6698\times10^{-13}$ & $2.6722\times10^{-13}\ [0.09\%]$ & $2.6730\times10^{-13}\	[0.12\%]$ & $2.7029\times10^{-13}\ [1.24\%]$ & $2.6777\times10^{-13}\ [0.30\%]$ \\
& $6$ & $2.5109\times10^{-11}$ & $2.5127\times10^{-11}\ [0.07\%]$ & $2.5133\times10^{-11}\	[0.09\%]$ & $2.5332\times10^{-11}\ [0.89\%]$ & $2.5186\times10^{-11}\ [0.31\%]$ \\
& $7$ &	$1.0972\times10^{-10}$ & $1.0980\times10^{-10}\ [0.08\%]$ & $1.0982\times10^{-10}\	[0.09\%]$ & $1.1044\times10^{-10}\ [0.66\%]$ & $1.0979\times10^{-10}\ [0.07\%]$ \\
& $8$ &	$8.8366\times10^{-09}$ & $8.8428\times10^{-09}\ [0.07\%]$ & $8.8439\times10^{-09}\	[0.08\%]$ & $8.8832\times10^{-09}\ [0.53\%]$ & $8.8253\times10^{-09}\ [0.13\%]$ \\
\hline
\end{tabular}
\end{center}
\caption{Energy flux for a circular orbit at radius $R=7.9456M$ around a nonrotating BH. We compare the results from the semianalytical method (subscript $ODE$), the full numerical scheme, extrapolated to null infinity (subscript $PDE_{\mathcal{I}^+}$) and extracted at finite radii, $R=1500M$ and $R=250M$ (subscripts $PDE_{1500}$ and $PDE_{250}$), and the results of \cite{N11} (subscript $Bernuzzi+11'$), which were calculated at null infinity. The relative differences, with respect to the semianalytical result, appear in square brackets.}
\label{table:d1}
\end{table*}
\endgroup
\subsection{Expansion at the Newtonian limit}
Equation (\ref{eq:d7}) can be analytically solved in the Newtonian limit, where it becomes equivalent to Eq. (\ref{eq:ext1}), with a specific value of $\omega=m\Omega$. Therefore, the solution for $f(r)$ is a combination of Bessel functions of the first and second kind, as in Eq. (\ref{eq:ext2}). Imposing outgoing wave boundary conditions and using the known limits of the Bessel functions yield\footnote{At the region $r<R$, all we need for this calculation is the asymptotic behavior $f|_{r\rg R^-}\propto r^{\li+1}$.}:
\begin{equation} \label{eq:aB11}
f(r>R)=\beta\left\{\def\arraystretch{1.5}\begin{tabular}{@{}l@{\quad}l@{}}
  $\frac{\Gamma\left(\li+\frac{1}{2}\right)}{\sqrt{\pi}}\lp\frac{2i}{m\Omega r}\rp^\li$ & $r\sim R$ \\
  $e^{i m\Omega r}$ & $r\gg R$
\end{tabular}\right.,
\end{equation}
where $\Gamma$ is the gamma function. $\beta$ is uniquely determined by imposing the discontinuity conditions at $r=R$:
\begin{equation} \label{eq:aB111}
\left\{\def\arraystretch{1}\begin{tabular}{@{}l@{\quad}l@{}}
  $\Delta f^{(\lambda)}|_{R}=\gamma^{(\lambda)}$\\
  $\Delta f^{\prime(\lambda)}|_{R}=-\frac{\delta^{(\lambda)}\gamma^{(\lambda)}}{R}$
\end{tabular}\right.,
\end{equation}
where $\gamma^{(\lambda)}=\frac{16\pi}{\Lambda(\Lambda-2)}\left\{\def\arraystretch{0.5}\begin{tabular}{@{}l@{\quad}l@{}}
  $Y^*_{\li m}(\frac{\pi}{2},0)$ & even\\
  $-\partial_\Theta Y^*_{\li m}(\Theta,0)|_{\Theta=\frac{\pi}{2}}/\sqrt{R}$ & odd
\end{tabular}\right.$ and $\delta^{(\lambda)}=\left\{\def\arraystretch{0.5}\begin{tabular}{@{}l@{\quad}l@{}}
  $\Lambda/2$ & even\\
  $1$ & odd
\end{tabular}\right.$. 

The radiated fluxes can be calculated using Eq. (\ref{eq:c1}) and (\ref{eq:c3}). For the energy flux, we get
\begin{equation} \label{eq:aB12}
\begin{aligned}
     \dot{E}_{\ell,m} =&\frac{2m^{2(\ell+1)}(2\ell+1)(\ell+1)(\ell+2)}{R^{\ell+3}\ell(\ell-1)(2\ell+1)!!^2} \\
    &\ \times\left\{\def\arraystretch{1}\begin{tabular}{@{}l@{\quad}l@{}}
         $\frac{(\ell-m)!(\ell+m)!}{(\ell-m)!!^2(\ell+m)!!^2}$ & $(\li+m)$ even  \\
         $\frac{4}{R}\frac{(\ell-m)!(\ell+m)!}{(\ell+1)^2(\ell-m-1)!!^2(\ell+m-1)!!^2}$ & $(\li+m)$ odd
     \end{tabular}\right..
\end{aligned}
\end{equation}
The result for the even energy flux is in accordance with a previous calculation of \cite{P93a}. The total flux, for a given multipole $\li$, can be evaluated up to subleading order as $\dot{E}_{\ell}=\sum_m \dot{E}_{\ell,m}\approx \dot{E}_{\li,\li}+\dot{E}_{\li,\li-2}$, where the contribution of the odd multipoles is negligible as $\dot{E}^{(o)}_{\ell}/\dot{E}^{(e)}_{\ell}\approx O\big(\frac{1}{R\ell}\big)$. From Eq. (\ref{eq:aB12}), we get that asymptotically for high multipoles, $\li\gg1$:
\begin{equation}
\begin{aligned}
    &\dot{E}_{\ell}  \approx
         \frac{1}{2R^3}\sqrt{\frac{\li}{\pi}}\left(\frac{e^2/4}{R}\right)^\ell\mathcal{B}_E(\li)\\
     &\mathcal{B}_{E}(\li) \sim 1.01+\frac{3.42}{\li}+\frac{4.08}{\li^2}
\end{aligned}
\end{equation}
The linear momentum flux is given by
\begin{equation}
\begin{aligned}
    \left|\dot{P}_{\li,m}\right|=&\frac{2}{R^{\li+7/2}}\frac{\left(\li+2\right)}{\li(\li+1)}m^{2(\li+1)}\Big(h^{(+)}_{\li,m}+h^{(-)}_{\li,m}\Big)\\
    &\times\frac{(\li-m)!(\li+m)!}{(\li-m)!!^2(\li+m)!!^2(2\li+1)!!^2},
\end{aligned}
\end{equation}
where, as in the energy flux case, the contribution of the odd multipoles is negligible, and we denote
\begin{equation}
\begin{aligned}
    h^{(\pm)}_{\li,m}=&\left(1\pm\frac{1}{m}\right)^{\li+1}\left(\li\pm m+1\right)\\
    &\times\left[\frac{(m\pm1)(\li+3)}{2\li+3}\mp4(\li\mp m) \frac{2\li+1}{\li(\li-1)}\right].
\end{aligned}
\end{equation}
For $\li\gg1$,
\begin{equation}
    \begin{aligned}
         &\left|\dot{P}_{\ell}\right| \approx
         \frac{e}{4R^3}\sqrt{\frac{\li}{\pi R}}\left(\frac{e^2/4}{R}\right)^\ell\mathcal{B}_P(\li) \\
         &\mathcal{B}_P(\li) \sim 1.001+\frac{3.515}{\li}+\frac{2.695}{\li^2}.
    \end{aligned}
\end{equation}

We note that $\dot{P}_{\li,m}\propto e^{i\Omega(t-r*)}$, and therefore, it vanishes when averaging over one period of the motion. Thus, as can be inferred from the symmetry of this case, there is no accumulation of recoil velocity in circular orbits.
\section{MULTIPOLAR DECOMPOSITION}
We present in Table \ref{table:f1} the detailed numerical results for the accumulated recoil velocity.
\begingroup
\begin{table}[ht]
\centering
\begin{tabular}{|c c || c c || c c | } 
\hline
$\li$ & $\left(V/\nu^2\right)$ & $\li$ & $\left(V/\nu^2\right)$ & $\li$ & $\left(V/\nu^2\right)$ \\
\hline
$2$ & $0.0289$ & $5$ & $0.0436$ & $8$ & $0.0460$ \\
\hline
$3$ & $0.0374$ & $6$ & $0.0448$ & $9$ & $0.0463$\\
\hline
$4$ & $0.0414$ & $7$ & $0.0456$ &  $10$ & $0.0464$\\
\hline
\end{tabular}
\caption{The accumulated recoil velocity up to a given multipole $\li$.}
\label{table:f1}
\end{table}
\endgroup
\bibliographystyle{unsrt}
\bibliography{main}

\end{document}